\documentstyle[aps,multicol,psfig,epsf,epsfig,axodraw]{revtex}
\begin{document}
\def\pa{\parallel}
\def\pe{\bot}
\let\a=\alpha \let\b=\beta  \let\c=\chi \let\d=\delta  \let\e=\varepsilon
\let\f=\varphi \let\g=\gamma \let\h=\eta \let\k=\kappa  \let\l=\lambda
\let\m=\mu   \let\n=\nu   \let\o=\omega    \let\p=\pi
\let\r=\varrho  \let\s=\sigma \let\t=\tau   \let\th=\vartheta
\let\y=\upsilon \let\x=\xi \let\z=\zeta
\let\D=\Delta \let\F=\Phi  \let\G=\Gamma  \let\L=\Lambda \let\Th=\Theta
\let\O=\Omega
\newcommand{\ie}{\hbox{\it i.e.\ }}
\def\xv{ {\bf x}}
\draft
\tightenlines
\title{Some puzzling problems in nonequilibrium field theories}
\author{
 Miguel A. Mu\~noz}  
\address{
Instituto de F{\'\i}sica Te\'orica y Computacional Carlos I,
Universidad de Granada, Facultad de Ciencias, 18071-Granada, Spain. \\}
\date{\today}
\maketitle
\begin{abstract}
I review some open problems on the ever-growing field of non-equilibrium 
phase transitions, paying  special attention to the formulation of 
such problems in terms of Langevin equations or, equivalently, 
field-theoretical descriptions, and their solution using renormalization 
group techniques.
\end{abstract}
%\pacs{PACS numbers:}
\begin{multicols}{2}
\narrowtext
\date{\today}  

\section{Introduction}
 The introduction of the Renormalization Group (RG) ideas and 
their application to the study of critical phenomena 
constitutes one of the milestones in the spectacular
 development of the Statistical Physics during the last quarter
 of the twentieth century. 
The RG proved to be
not only  a powerful analytical tool to deal with complex problems (\ie,
problems with many different relevant scales), 
but also a conceptually beautiful and elegant theory, with
a huge range of applicability.   
While RG ideas applied in (discrete) real space have helped to shed light on
many problems \cite{RS}, it has been in the framework of (continuous) field
theoretical descriptions where, 
combined with perturbative methods, the RG has reached its 
most successful expressions. 
In particular, given a continuous field theoretical 
representation of a given statistical system at thermodynamical equilibrium, 
the identification of its critical points with fixed points of a
conveniently defined RG transformation permits to obtain 
(perturbative) analytical expressions
for the corresponding critical exponents. 
At the same time, by allowing to distinguish 
 {\it relevant} from {\it irrelevant} ingredients
in a rather systematic way,
the combined
use of field theories and the RG, has permitted to obtain elegant classifications of 
equilibrium critical phenomena and put under firm basis the concept of
 {\it universality} \cite{ZJ,RS}. 
The representation of the Ising  model universality 
class by the $\lambda \phi^4$ theory and its subsequent 
renormalization constitutes a paradigmatical instance \cite{ZJ}.

Given this rather satisfactory scenario, theoreticians started wondering whether
 also critical phenomena occurring in systems away from equilibrium 
could be attacked using similar tools.
In the lack of a well established theory for non-equilibrium phenomena, it is not 
straightforward to extend the equilibrium formalism to deal with non-equilibrium 
problems (for example, in these cases there is no partition function 
to be renormalized).
 The best way to do so turned out to be expressing such  problems
in terms of Langevin equations describing the underlying {\it dynamics} at
a (continuous) coarse grained level. This procedure is valid not only to study
general non-equilibrium processes but also relaxation to equilibrium states 
(the model A and B for the non-conserved and conserved  relaxation
dynamics of the Ising class  are prototypical examples).  
In some cases Langevin equations representing given microscopic processes
 can be derived analytically using different techniques
(among them: Fock space formalism combined with coherent state 
representations \cite{Fock}, and Poisson transformations \cite{Gardiner}), 
while in  many other cases
they are just postulated from a phenomenological ground, by respecting 
what are considered {\it a priori} to be main symmetries, 
conservation laws, and other relevant dynamical constraints. 

 Experience teaches us that the richness and variety of phase 
transitions occurring away from equilibrium is by far much larger than
 that of equilibrium, and that in many cases it is very difficult to decide
 {\it a priori} what are the essential ingredients to be put     
into a sound Langevin description. Therefore, developing rigorous and systematic
techniques envisaged to derive coarse grained Langevin equations from 
microscopic models is a high priority task within this context. 
On the lack of such general approaches one has to rely many times 
on phenomenological approaches.

 Before proceeding, let us remark that any arbitrary Langevin 
equation can be written as an equivalent Fokker-Planck equation \cite{vK,Gardiner}, 
and its solution expressed in terms of a generating functional 
(or equivalently  and effective {\it action}) obtained as a path integral 
representation of the stochastic Langevin process.  
Therefore, in what follows ``Langevin equations'', 
``Fokker-Planck equations'', ``generating functionals'', or
 ``field theoretical actions'' (Liouville 
operator) will be used interchangeably (see \cite{BJW,ZJ,Cardy,Fock,Gardiner}).

 In the forthcoming sections we report on a variety of interesting
non-equilibrium systems, and present a {\it list of open problems} within 
this field.

  \section{The driven Lattice gas (DLG)}

The DLG is a variation of the kinetic Ising model with conserved dynamics,
in which transitions in the direction (against the direction)
of an externally applied field, $\vec{E}$, are favored
(unfavored) \cite{MD,katz,zia}, while transitions perpendicular
to the field are unaffected by it.
The external field induces two main non-equilibrium effects:
(i) the presence of a net current of particles along its direction,
and (ii) {\it strong} anisotropy.
At high temperatures, the system is in a disordered phase while
lowering the temperature there is 
(for half-filled lattices) 
a continuous transition into an ordered 
phase with high and low density aligned-with-the-field stripes.
Elucidating the DLG critical properties
is an important issue in the way to rationalize the
behavior of non-equilibrium systems.
The following Langevin equation was proposed some years back based on 
phenomenological arguments \cite{JS,zia}:
\begin{eqnarray}
\label{dds}
\partial_t\phi({\bf r},t)= &&
\tau_\pe \nabla_{\pe}^2\phi
-\nabla_{\pe}^4\phi
+{ \lambda \over6}\nabla^2_{\pe}\phi^3
 \nonumber \\
&& + \tau_\pa  \nabla_{\pa}^2 \phi
- \alpha  \nabla_{\pa} \phi^2
+ {\bf \eta}({\bf r},t),
\label{DDS}
\end{eqnarray}
where $\phi$ is the coarse grained field, ${\bf \eta}$ is a conserved
Gaussian noise, and where the cubic term (a dangerously irrelevant
variable \cite{ZJ}) is kept in order to ensure stability \cite{JS}.
The fact that some of the predictions derived analytically from 
this equation (for instance, the order parameter critical exponent
 $\beta$ takes a value $1/2$) have not been convincingly 
verified numerically (a value $\beta \approx 0.33$ is systematically found
in 2-dimensional Monte Carlo simulations of the DLG and variations of it 
\cite{Fattah,MD,Albano}), has triggered further studies. 
These new analyses ended up with the proposal of a new Langevin equation aimed 
at describing the critical properties of the DLG:
\begin{equation}
\label{ads}
\partial_t\phi({\bf r},t)=
\tau_\pe \nabla_{\pe}^2\phi
-\nabla_{\pe}^4\phi
+{\lambda \over6}\nabla^2_{\pe}\phi^3
 + \tau_\pa \nabla_{\pa}^2 \phi
+ {\bf \eta}.
\label{ADS}
\end{equation}
This equation
is a well known one: it coincides with the Langevin
equation representing the random DLG
(RDLG) \cite{RDDS} (for which
the driving field takes values $\infty$ and $-\infty$
in a random unbiased fashion, generating anisotropy
but not an overall current).
This equation has been extensively studied
in \cite{RDDS}; its critical dimension is $d_c=3$
(instead, $d_c=5$ for (\ref{DDS})) and its associated critical exponents
and finite size scaling properties are now well known.
At least two different recent numerical studies show that this
equation reproduces rather nicely all DLG critical properties, and support
the conclusion that it  is the anisotropy  and not the overall current  the
main ingredient characterizing the DLG phase transition.

 However, the situation is far from satisfactory. The central issue
is that naive power counting analysis shows that the current term 
establishing the difference between the two abovementioned theoretical 
descriptions is a relevant perturbation at the Eq.(\ref{ADS}) RG fixed point and 
therefore, it is unclear why it should be absent in the proper Langevin 
description. It has been argued in \cite{Nos,Fattah} that the
coefficient of this term 
happens to vanish in the limit $E \rightarrow \infty$. This would 
imply that the DLG has a sort of multicritical point in the infinite fast 
driving limit. This scenario still needs to be confirmed numerically 
\cite{Fattah,Albano}. 

 Another theoretical way out of this puzzling situation
 is that the non-linear current term should be absent 
due to the fact that the microscopic theory is {\it fermionic} (i.e. occupation 
number restricted to be $0$ or $1$), while the Langevin equation describes, 
in principle, a bosonic process: the current term is required in this 
bosonic formalism in order to have a vanishing current in perfectly 
ordered aligned-with-the-field stripes 
(for which, the fermionic restriction precludes the presence of a 
non-vanishing current). 
We are presently working in the derivation of a 
field theoretical description that takes properly into account the 
fermionic nature of the microscopic model.  
 
\section{Systems with many absorbing states}
Maybe the best well-known genuine non-equilibrium Langevin equation is the,
so-called, {\it Reggeon} field theory \cite{RFT}
\begin{equation}
  {\partial_t \rho ( {\bf x}, t )}
 = \nabla \rho ( {\bf x}, t )+ a \rho ( {\bf x}, t )
-b \rho ^ 2( {\bf x}, t )
 + \sqrt{\rho} \eta ( {\bf x}, t )
\label{RFT}
\end{equation}
that captures the critical properties of phase transitions into a
single absorbing state (with no extra symmetries nor conservation laws),
usually referred to as the directed percolation (DP) universality class 
\cite{MD,Granada,Hin}.
The key property of  Eq.(\ref{RFT}) is that all terms (including
the noise) vanish in the absence of activity, \ie for $\rho({\bf x})=0$). 
 Even though convincing experimental realizations of this broad 
universality class are still missing,
an overwhelming number of models have been studied, all of them 
sharing their critical behavior with this minimal Langevin equation.
 The situation is less satisfactory for systems with many 
different absorbing states \cite{Muchos},
a prototype of which is the Pair Contact Process (PCP)
\cite{PCP,IAS}. In the PCP pairs of particles can generate new
particles or get annihilated, but isolated particles do not have any dynamics;
 any configuration with just isolated particles is therefore absorbing. 
Models of this are relevant, for example, in catalysis
\cite{Muchos}.
The following Langevin equation for the PCP and related models was proposed some 
years back:
\begin{eqnarray}
& {\partial_t \rho ( {\bf x}, t )}
= D \nabla^2 \rho ( {\bf x}, t )+ a \rho ( {\bf x}, t )
-b \rho ^ 2( {\bf x}, t )  \nonumber\\
& + \alpha \rho ( {\bf x}, t )e^{-w_1 \int_0^t
\rho ({\bf x} ,s) ds } + \sqrt{\rho} \eta ( {\bf x}, t )~
\label{Lan2}
\end{eqnarray}
where the field $\rho$ in Eq.(\ref{RFT}) represents the density of
pairs (activity), and the effect of the isolated particles (characterizing the
different absorbing states 
\cite{IAS}) is captured in the non-Markovian exponential term.
 It has been argued that the critical properties of this
equation when approaching the critical point from the active phase are
DP like \cite{IAS,slave}. Indeed, it is straightforward to see that in the presence
of non-vanishing  stationary activity the exponential term cancels out and we are
left simply with DP. 
On the contrary, for spreading experiments for which critical propagation
can occur inside the absorbing phase, the exponential term can be expanded
in power series, and one ends up with \cite{IAS}
\begin{eqnarray}
& { \partial_t \rho ( {\bf x}, t )}
= D_2 \nabla^2_{\bf x} \rho ( {\bf x}, t )+ a \rho ( {\bf x}, t ) 
\nonumber \\ &
+ \alpha \rho ( {\bf x}, t ) \int_0^t
\rho ({\bf x} ,s) ds  + \sqrt{\rho} \eta ( {\bf x}, t )~
\label{dynp}
\end{eqnarray}
which  is the well-known Langevin equation describing isotropic 
percolation dynamically \ie {\it dynamical percolation} (DyP) \cite{dyp}.
These results are rather satisfactorily reproduced in numerical 
simulations \cite{IAS}. Still there is a point which remains obscure:
 If one works right at the critical point of the full theory, 
dynamical percolation terms are generated perturbatively, as first observed
in \cite{IAS} and recently stressed in \cite{Fred}. 
This new vertex being more relevant than the dominant 
non-linearity in Eq. (\ref{RFT}), leaded van Wijland to propose that 
the true asymptotic critical behavior should be controlled by a DyP fixed point. 
In order to generate an active phase (missing in DyP) he proposes to treat
the term proportional to $n^2$ as a {\it dangerously irrelevant operator}, 
and finds an analytical expression for $\beta$. 
We believe that such a calculation cannot 
apply  to the PCP since,
even including the new term lacks of a well defined 
active phase. Being more precise, a term $-b \rho$ cannot compensate 
the linear in time divergence of $ \alpha \rho ( {\bf x}, t ) \int_0^t 
 \rho ({\bf x} ,s) ds $  in the active phase.

Another open problem in this context is
understanding within a systematic RG calculation how the background field
(describing  the different absorbing configurations)
emerges as a {\it slave mode} of the activity field, \ie how it inherits 
the critical properties of the order parameter in the active phase \cite{slave}.
A comprehensive understanding of this 
family of phase transitions putting together the active and 
inactive phases is still missing. 

\section{Self-organized criticality}
 The observation that  {\it sandpiles}, the archetype of 
self-organized systems \cite{soc}, fall into different absorbing states 
after every avalanche, right before new sand is added, opened 
the door to rationalize their critical properties using Langevin 
equations similar to those described in the preceeding section.
The first step in order to do so was to regularize the sandpiles, by
 introducing the, so-called, fixed energy
sandpiles (FES) which eliminate dissipation and addition of energy (sand-grains)
\cite{FES}.
This converts the total energy into a control parameter: 
large total energy generates stationary activity, 
while small amounts of energy lead the system with certainty to an
absorbing  configuration. The proposed set of equations for FES are:
\begin{eqnarray}
\partial_t \rho(\xv,t) &= & a \rho (\xv,t) - b \rho^2(\xv,t)
 + \nabla^2 \rho(\xv,t) \nonumber \\
 & +  & w E(\xv,t) \rho(\xv,t)+
 \sqrt{\rho} \eta(\xv,t)
\nonumber \\
\partial_t E(\xv,t) &= & \lambda \nabla^2 \rho(\xv,t)
\label{Lan}
\end{eqnarray}
where $a, b, w$, and $\lambda$ are constants, and
$\eta$ is a Gaussian white noise.
In these equations the activity dynamics is controlled by the same type
of terms appearing in Eq.(\ref{RFT}), plus an
additional coupling between the activity
field and a static {\it conserved} energy field. This extra
term stems from the fact that creation of activity
is locally fostered by the presence of a high background field density,
 and the energy is a conserved field, $ E(\xv,t)$.
The extra conservation law is therefore 
a new (relevant) ingredient with respect to RFT.
Some other terms, consistent with the
symmetries and conservation laws, could have been included
in Eq.(\ref{Lan}) but they all turn out to be irrelevant from a
power counting analysis \cite{FES}. 
This same set of Langevin equations (plus higher order noise  terms)
 has been derived using Fock-space
techniques for other discrete models with many absorbing states and
a static local conservation law \cite{romu}.
The field theoretical analysis of this set of equations turns out to be a
delicate issue (observe that analogous field theories for models with absorbing
states but where the conserved field is not a static one 
can be studied perturbatively without any problem, \cite{WOH}).
 As happens in the case of many absorbing states (without 
a conservation law) also here, at criticality DyP type of terms are generated.
Here, even the physics coming from the active phase is not easy to work out.
In this context, it has also been recently proposed \cite{Fred} 
that the critical properties should be
described by the ``regularized'' DyP fixed point, and 
again similar criticisms as those
made before could apply here (although in this case, the problem is even 
more involved).

A successful RG calculation of this theory would be extremely valuable from a
theoretical perspective; it would not only determine the critical exponents
for a vast class of {\it self-organized} systems, but also clarify the issue
of the proposed connection between self-organized criticality and the
pinning of {\it interfaces in disordered media} \cite{Alava}.

Another related problem is that of the study of the effect of
quenched disorder in systems with absorbing states \cite{FES}.
 A field theoretical analysis by Janssen \cite{Quenched}, 
revealed the existence of running away RG trajectories, 
whose correspondence with the observed phenomenology in 
$d=1$ and $d=2$ \cite{Quenched2} remains mysterious.

Before finishing this section, we want to point out that 
a promising formulation of the same problem, namely deriving an effective action
for sandpiles has been recently addressed in \cite{Vidigal}.

\section{Other Reaction Diffusion Systems}
 In this section we briefly enumerate some other open problems in 
field theoretical analyses of general reaction
diffusion processes.

%\subsection{Complex noise}
%Reaction-limited annihilation processes as $A+A \rightarrow 0$ can be easily cast
%into an effective action using a Fock-space techniques combined with a coherent
%state representation. In terms of a Langevin equation, one gets
%\begin{equation}
%\partial_t \rho(\xv,t) =   - 2  \rho^2(\xv,t)
% + \nabla^2 \rho(\xv,t) + i \sqrt{2}  \sigma \rho(\xv,t) \eta(\xv,t)
%\label{complex}
%\end{equation}
%where the noise as can be seen is purely {\it imaginary}. 

\subsection{Two symmetric absorbing states}
For some time it was believed that parity conservation (PC) was the main ingredient 
of a new, non DP, universality class \cite{baw}. 
By now, it is well established that the presence of an exact $Z_2$ 
symmetry between to equivalent absorbing states 
is its main distinctive trait \cite{hin}. 
Also, the introduction of parity conservation has been shown to play no 
relevant role in reaction-diffusion binary spreading
models (\cite{hin}, see also \cite{hug}). 
A field theoretical description of this universality 
class was proposed by Cardy and T{\"a}uber some years back.
It starts with a Fock-space representation of the 
reaction diffusion lattice model in this class:
$A+A \rightarrow 0$, $ A\rightarrow (m+1) A$ (with $m$ an even constant)
 allowing to derive a field theoretical action. 
Even though the analysis of such a theory
(that guarantees that the parity in
the number of particles is conserved)
is based on some uncontrolled expansion, it reproduces nicely many general 
features of this family, including the existence of a non-trivial critical
 point below two dimensions, the critical dimension $d_c=2$,
 and some other properties. 
The formalism is also applicable to odd values of $m$ where it also 
generates sound results \cite{tauber}. 

An interesting open problem in this perspective would be to construct a 
more adequate field theoretical representation that should include as
the main ingredient the presence of two symmetric $Z_2$ absorbing states,
in the hope that in this more natural language a detailed standard
RG procedure would be applicable.

 Another related problem is that of writing  down and analyzing a
field theory for the Voter model \cite{Voter} and extensions of it,
for which a similar symmetry between different absorbing states 
appears (also the non-equilibrium kinetic Ising model
at zero temperature belongs two this family of models with $Z_2$-symmetric
absorbing states \cite{baw}).
 
\subsection{Pair Contact Process with diffusion (PCPD)}
 In recent years, a new single-component absorbing-state
 universality class has been unveiled. It is the so called PCPD: if
in the standard PCP we allow for diffusion of isolated particles we
are led to this new class. Observe that switching-on diffusion 
represents a singular perturbation as, for instance, the many PCP  absorbing 
configurations are reduced just to two, an empty one and one with 
a single wandering particle.  It seems that the main ingredient in this class
is the fact that reactions are binary (two particles are required for
reactions to occur) and solitary particles travel performing random walks
in between reaction zones. A field theory for this model was worked out
by Howard and T\"auber \cite{howard} some years back. Using  a bosonic field theory 
(exact for a version of the model without a stationary active phase,
usually called annihilation-fission process)
they concluded that the critical dimension is $d_c=2$ 
 and that the transition is not DP-like. 
Unfortunately, the theory turned out to be 
non-renormalizable (i.e. an infinite hierarchy of relevant operators 
are generated perturbatively, making it un-tractable).

Different proposals have been made recently in order to rationalize
 the critical behavior of this class. These go from
 the existence of continuously variant
exponents (as a function of the diffusion constant), to the existence of
two universality classes (for small and large diffusion constants 
respectively), or just one well-defined set of critical exponents 
\cite{PCPD}. 

 One possible strategy to analyze this problem from a field theoretical
 point of view is to introduce discrete models in this
class with two different species: one corresponding to diffusing ``isolated''
particles, and one ``diffusing-reacting'' type of particle playing the role
of the pairs in the original model \cite{hh,slave}. This leads to a set
of Langevin equations analogous to those proposed for the 
PCP, Eq.(\ref{Lan2}) but including diffusion of the secondary field.
 This changes the critical dimension from $d_c=4$ to $d_c=2$, 
but  a systematic perturbative analysis allowing for a determination 
of the critical exponents has not been completed so far.

Similar problems are observed upon studying ternary-reactions
(as $3A \rightarrow 0$, combined with  $3A \rightarrow 3+m$)
for which  new critical behavior is expected \cite{triplet,hug}. 
For higher order nth-reactions the upper
critical dimension for annihilation is below $d=1$ therefore no
anomalous phase transition is expected to occur \cite{tauber,hug}.

\section{Discussion}

We have briefly reviewed some non-equilibrium field theoretical 
open problems.
They are related to non-equilibrium Ising-like models as well as systems
with absorbing states. Other families of problems not discussed here are
interfacial growth, non-equilibrium wetting phenomena, transitions described by
the multiplicative noise equation as for example those occurring in the 
synchronization of coupled-map-lattices, etc.
The existence of the various problems reported here 
gives raise to the following priorities
for the developing of a systematic non-equilibrium field theoretical formalism:
 
i) the necessity of developing new tools for deriving Langevin equations
(or field theories) in a systematic, rigorous way, 
from discrete microscopic models.

ii) Understanding the role of hard-core repulsion and/or implementing
this constraint in a systematic way in field theoretical descriptions
\cite{fermionic}.

iii)  Developing new analytical schemes, specially in low dimensions, 
to deal with problems for which standard epsilon-expansion does not yield 
satisfactory results. 

  It is my hope that this brief overview will stimulate further studies in 
this field.

\vspace{1cm}

%%%%%%%%%%%%%%%%%%  ACKNOWLEDGMENTS  %%%%%%%%%%%%%%%%%%%%%%%%%%%
It is a pleasure to acknowledge 
A. Achahbar, H. Chat\'e, C. da Silva Santos, R. Dickman, P. L. Garrido,
G. Grinstein,  R. Livi,  J. Marro, R. Pastor-Satorras,
M. A. Santos, Y. Tu, F. van Wijland, A. Vespignani, and
S. Zapperi, for very enjoyable collaborations and/or enlightening 
discussions on the issues discussed in this paper. 
I acknowledge financial support from the European Network contract 
ERBFMRXCT980183, and the Spanish Ministerio de Ciencia 
y Tecnolog{\'\i}a (FEDER), under project BFM2001-2841.

%%%%%%%%%%%%%%%%%%%%  REFERENCES %%%%%%%%%%%%%%%%%%%%%%%%%%%%%%%%%

\end{multicols}
 
\end{document}